\documentclass[12pt]{iopart}

\usepackage{graphicx} 
\usepackage{subfigure}
\begin{document}

\title{Are eccentricity fluctuations able to explain the centrality dependence of $v_4$?}

\author{\underline{Cl\'ement Gombeaud} and Jean-Yves Ollitrault}

\address{Institut de physique th\'eorique
CEA-Saclay
91191 Gif-sur-Yvette, France}
\ead{clement.gombeaud@cea.fr}
\begin{abstract}

The fourth harmonic of the azimuthal distribution of particles $v_4$ has been measured for Au-Au collisions at the Relativistic Heavy Ion Collider (RHIC). The centrality dependence of $v_4$ does not agree with the prediction from hydrodynamics. In particular, the ratio
$v_4/(v_2)^2$, where $v_2$ denotes the second harmonic of the azimuthal distribution of particles, is significantly larger than predicted by hydrodynamics.
We argue that this discrepancy is mostly due to elliptic flow ($v_2$) fluctuations. We evaluate these fluctuations on the basis of a Monte Carlo Glauber calculation. The effect of deviations from local thermal equilibrium is also studied, but appears to be only a small correction.
Combining these two effects allows us to reproduce experimental data for peripheral and midcentral collisions. However, we are unable to explain the large magnitude of $v_4/(v_2)^2$ observed for the most central collisions.


\end{abstract}

\maketitle

\begin{section}{Introduction}

The azimuthal distribution of emitted particles is a good tool for understanding the 
bulk properties of the matter created in non central nucleus-nucleus collisions. 
In the center of mass rapidity region, it can be expanded in Fourier series:
\begin{equation}
\frac{dN}{d\phi} \propto 1+2 v_2 \cos(2\phi)+2 v_4 \cos(4\phi)+\cdots
\end{equation}
where $\phi$ is the azimuthal angle with respect to the direction of the impact parameter, and odd 
harmonics are zero by symmetry.
The large magnitude of elliptic flow $v_2$ observed at RHIC suggests that the matter
created in Au-Au collisions behaves like an almost perfect fluid. However, recent 
experiments~\cite{Abelev:2007qg,Huang:2008vd} observe that, at midrapidity and fixed $p_t$,  $v_4 \simeq (v_2)^2$,
while ideal hydrodynamics predicts that 
$v_4=\frac{1}{2}(v_2)^2$~\cite{Borghini:2005kd}.
In this talk, I investigate this discrepancy.
\end{section}

\begin{section}{Fluctuations in initial conditions}
  \begin{figure}[!ht]
\centering
\includegraphics[width=2.4in]{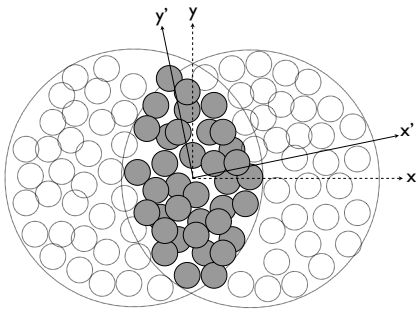}
  \caption{(Color online) Picture of the two frames used for defining the
  initial eccentricity (from~\cite{Alver:2008zz}). The $x$ axis defines the
  reaction plane while the $x'$ axis is the minor axis of the ellipse
  drawn by the participating nucleons (grey dots). }
  \label{fig:fig1}
\end{figure}

Figure~\ref{fig:fig1} presents a schematic picture of a 
non central heavy-ion collision (HIC). The overlap area of 
the nuclei has an almond shape, which generates elliptic flow.
 However, the matter is not continuously distributed in a nucleus. The
 positions of the nucleons in the colliding nucleus are important: they also draw
 an ellipse which differs from the overlap area both in eccentricity
 and in orientation. From one event to the other, even at fixed impact parameter,
  the positions of the nucleons in the nucleus fluctuate. The participant plane 
  eccentricity ($\epsilon_{PP}$), defined as the eccentricity of the ellipse drawn
  by the participating nucleons~\cite{Miller:2003kd, Bhalerao:2003xf}, thus fluctuates. Since elliptic flow appears to be driven by
  this participant plane eccentricity, these eccentricity fluctuations translate into fluctuations
  of the flow coefficients $v_2$ and $v_4$~\cite{Ollitrault:2009ie}.
  
\end{section}

\begin{section}{Modeling eccentricity fluctuations}

The initial distribution of energy, which is needed
 to compute the initial eccentricity in a HIC, is poorly known.
 In this talk I use a
 specific model, based on a Monte Carlo Glauber (MCG) calculation~\cite{PHOBOSMCG}.
 The initial eccentricity is given for each event by:
\begin{equation}
\epsilon_{PP}=\frac{\sqrt{(\sigma_y^2 -\sigma_x^2)^2+4\sigma_{xy}^2}}{\sigma_x^2+\sigma_y^2}
\end{equation}
 where $\sigma_x^2=\langle x^2 \rangle - \langle x \rangle^2$ and $\sigma_{xy}=\langle xy \rangle - \langle x \rangle \langle y \rangle$
 and the $\langle \rangle$ denote averages over participating nucleons.
 Each participant nucleon is given a weight proportional to the number of particles it creates, according to the two-component picture:
 $w=(1-x) + x N_{coll-nucleon}$ where $N_{coll-nucleon}$ is the number of binary collisions of the nucleon.
 The sum of weights scales like the multiplicity:
\begin{equation}
\frac{dN_{ch}}{d\eta}=n_{pp}\left[ (1-x)\frac{N_{part}}{2}+xN_{coll}\right ].
\end{equation}
 where $N_{part}$ and $N_{coll}$ are respectively the number of participants and of binary collisions of the considered event.
 We choose the value $x=0.13$ which best describes the charged hadron multiplicity observed experimentally~\cite{x13data}.
 We define the centrality according to the number of participants. We evaluate eccentricity fluctuations in centrality classes containing $5\%$ 
 of the total number of events. We do not introduce any hard core repulsion between nucleons in the MCG.
 
\end{section}

\begin{section}{How eccentricity fluctuations affect $v_4/v_2^2$.}
   \begin{figure}[!ht]
\centering
\includegraphics[width=2.7in, angle=270]{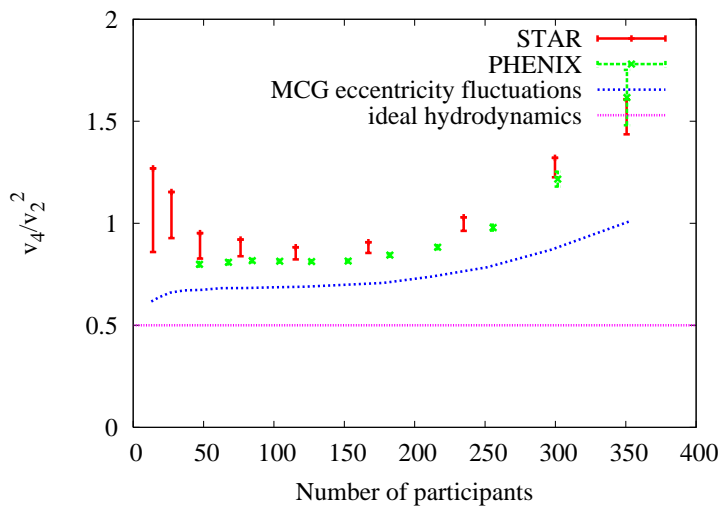}
  \caption{(Color online) Centrality
  dependence of $v_4/(v_2)^2$: data from 
  STAR~\cite{Y.Bai} and PHENIX~\cite{R.Lacey}; error bars on STAR data
  points are our estimates of nonflow
  errors~\cite{Gombeaud:2009ye}. Lines are  
  predictions from ideal hydro with or without fluctuations.} 
  \label{fig:fig2}
\end{figure}

There is no direct measure of the flow coefficients $v_2$ and $v_4$. 
They can be obtained using different analysis methods. The one I will consider now relies on 
azimuthal correlations between particles near midrapidity.
Experimentally, $v_2$ can be extracted from the 2-particle correlation and $v_4$ from the 3-particle 
correlation using $\langle \cos(2\phi_1-2\phi_2)\rangle=\langle (v_2)^2 \rangle$ and 
$\langle \cos(4\phi_1-2\phi_2-2\phi_3)\rangle=\langle v_4(v_2)^2 \rangle$, 
where angular brackets denote an average value within a
 centrality class. 
 Thus, any experimental measure of $v_4/v_2^2$ obtained using this method is rather a measure
 of $\langle v_4 \rangle / \langle v_2^2 \rangle^2$. Taking into account the ideal hydrodynamics prediction
 $v_4=\frac{1}{2}(v_2)^2$~\cite{Borghini:2005kd}, we obtain:
 \begin{equation}
 \left(\frac{v_4}{(v_2)^2}\right)_{\rm measured}=\frac{1}{2} \frac{\langle (v_2)^4 \rangle}{\langle
(v_2)^2 \rangle^2}>\frac{1}{2}.
\end{equation}
 Assuming that $v_2$ scales like the participant plane eccentricity $\epsilon_{PP}$, 
 the effects of fluctuations on $v_4/v_2^2$ is obtained by computing: 
  \begin{equation}
 \left(\frac{v_4}{(v_2)^2}\right)_{MCG} = \frac{1}{2} \frac{\langle \epsilon_{PP}^4 \rangle}{\langle \epsilon_{PP}^2 \rangle ^2}.
 \end{equation}
 
 The resulting prediction for $v_4/v_2^2$ is displayed in figure~\ref{fig:fig2}.
 Fluctuations clearly explain most of
the difference between hydro and data. It also appears that experimental data are
still slightly higher than our prediction from fluctuations. However, these predictions
are based on a specific parametrization of the initial conditions. 

\end{section}

\begin{section}{Flow fluctuations from experimental data}

  \begin{figure}[!ht]
\centering
\includegraphics[width=2.7in, angle=270]{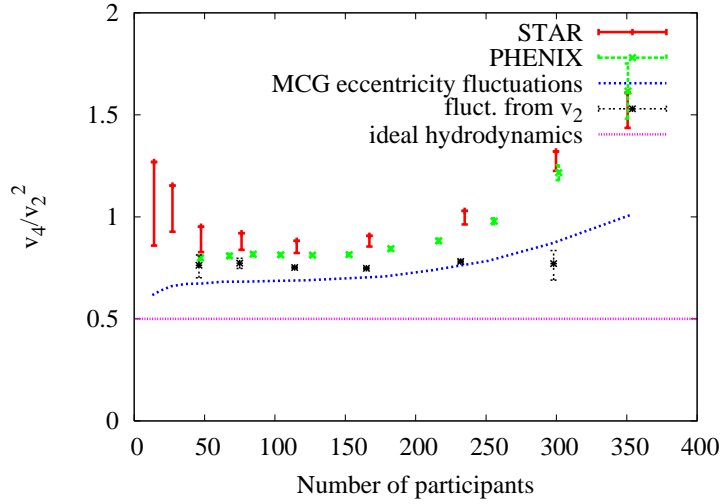}
  \caption{Same as figure \ref{fig:fig2}, additional points (labeled "fluct. from $v_2$")
  are obtained using equation~(\ref{v24value}) with $v_2\{4\}$ from \cite{refSTARv4} and $v_2\{2\}$ from \cite{refSTARv2}.} 
  \label{fig:fig3}
\end{figure}

Another possible way of evaluating flow fluctuations is to compare
the values of $v_2$ obtained using different analysis methods.
Elliptic flow can be obtained from both $2$-particle cumulants ($v_2\{2\}$) and from
$4$-particle cumulants ($v_2\{4\}$) using $v_2\{2\}^2=\langle (v_2)^2\rangle$ (neglecting the non-flow contribution) and
$v_2\{4\}^4=2\langle (v_2)^2\rangle^2-\langle (v_2)^4\rangle$.
Inverting the last equation leads to:
\begin{equation}
\label{v24value}
\left(\frac{v_4}{(v_2)^2}\right)=\frac{1}{2} \frac{\langle (v_2)^4 \rangle}{\langle(v_2)^2 \rangle^2}
=\frac{1}{2} \left (2-\left (\frac{v_2\{4\}}{v_2\{2\}}\right )^4\right).
\end{equation}

The values of $v_4/v_2^2$ obtained using this method are displayed on figure~\ref{fig:fig3}.
They overshoot slightly our results from MCG eccentricity fluctuations, but the overall agreement
remains good. This provides
a good check of our MCG prediction.
A small residual discrepancy remains between our prediction and the experimental data.
We argue that for peripheral to midcentral collisions, 
it may be understood in terms of deviations from local thermal
equilibrium. 
\end{section}

\begin{section}{Partial thermalization effects}
So far, I have only discussed how fluctuations
in initial conditions modify the prediction from ideal hydrodynamics
 for $v_4/v_2^2$.
 But ideal hydrodynamics relies on the very strong assumption that the 
 system remains in local thermal equilibrium (a regime where the average
  number of collisions per particle $n_{coll}$ is large) throughout the evolution. 
In a previous work~\cite{Drescher:2007cd} we have shown that, in order
to reproduce the centrality dependence of elliptic flow,
 the deviation from local thermal equilibrium must be taken into
 account ($n_{coll}\propto 3-5$ would be a typical value
 for Au-Au collisions at the top RHIC energy).
   \begin{figure}[!ht]
\centering
\includegraphics[width=3.7in]{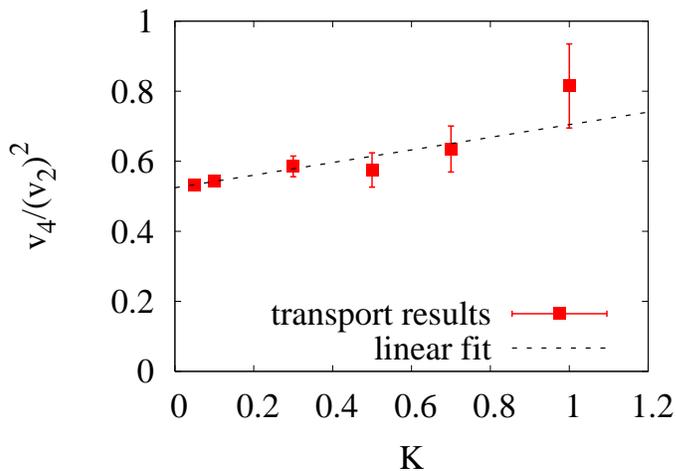}
  \caption{(Color online) Variation of $v_4/(v_2)^2$ with the Knudsen
  number.} 
  \label{fig:fig4}
\end{figure}

 Qualitatively, in the limit of small $n_{coll}$ (far from equilibrium),  one expects both $v_2$ and
$v_4$ to scale like $n_{coll}$, so that $v_4/(v_2)^2$ scales like
$1/n_{coll}$: we thus expect 
that the farther the system from equilibrium, the larger
$v_4/(v_2)^2$~\cite{Bhalerao:2005mm}.
In order to have a more quantitative estimate of the effects
of partial thermalization, we use a $2+1$-dimensional solution of the
 relativistic Boltzmann equation to study systems 
with arbitrary $n_{coll}$. We use the Knudsen
number~\cite{Bhalerao:2005mm}, $K\propto 1/n_{coll}$, as a measure of
the degree of thermalization of the system. 

Figure~\ref{fig:fig4} displays the dependence of $v_4/(v_2)^2$ with the Knudsen number. 
In the limit $K\rightarrow 0$, transport results show that $v_4/(v_2)^2=0.52$, which is close to $1/2$.
We also observe, as expected from the low $n_{coll}$ limit, that increasing $K$ leads to an increase of $v_4/(v_2)^2$.
But this effect is only a small correction.

  \begin{figure}[!ht]
\centering
\includegraphics[width=2.7in,angle=270]{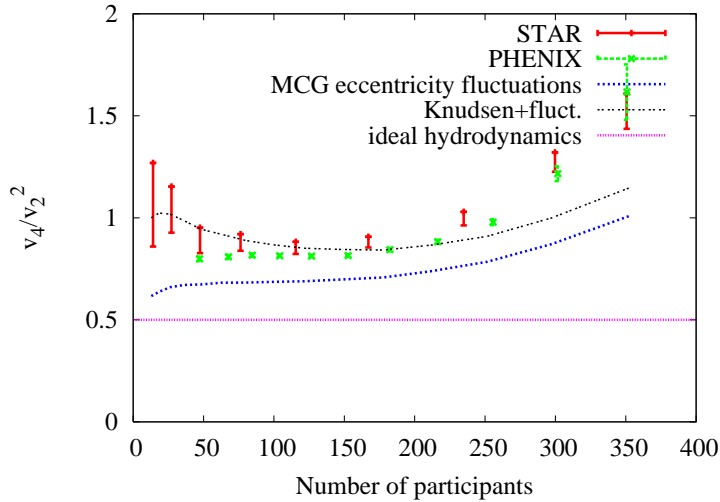}
  \caption{Same plot as figure~\ref{fig:fig2} (right), with one
  additional curve showing the effect of the deviation from local
  equilibrium.} 
  \label{fig:fig5}
\end{figure}

The effects of partial thermalization on the centrality dependence of $v _4/(v_2)^2$ are displayed on figure~\ref{fig:fig5}.
The values of the Knudsen number needed for this plot are borrowed from a previous study~\cite{Drescher:2007cd}.
Figure~\ref{fig:fig5} shows that adding the effects of deviation from local thermal equilibrium to the fluctuations, our prediction 
overshoots slightly the data for midcentral and
peripheral collisions, but the overall agreement is good. 
We do not yet
understand the large value of $v_4/(v_2)^2$ for central collisions.

\end{section}

\begin{section}{Conclusion}

To conclude, I would like to recall three points:
1) $v_4$ is mainly induced by $v_2$; 2)
the deviation from local equilibrium has a small effect on $v_4/(v_2)^2$;
3) eccentricity fluctuations explain the observed values of
$v_4/(v_2)^2$, except for the most central collisions which require
further investigation.
\end{section}

\section*{Acknowledgments} 
This work is funded by `Agence Nationale de la Recherche' under grant
ANR-08-BLAN-0093-01.\\
\newline


\end{document}